\documentstyle [12pt,psfig]{wspgam}
\newdimen\digitwidth
\setbox0=\hbox{\rm0}
\digitwidth=\wd0
\catcode`~=\active
\def~{\kern\digitwidth}
\input epsf

\def\spose#1{\hbox to 0pt{#1\hss}}
\def\lta{\mathrel{\spose{\lower 3pt\hbox{$\mathchar"218$}}
 \raise 2.0pt\hbox{$\mathchar"13C$}}}
\def\gta{\mathrel{\spose{\lower 3pt\hbox{$\mathchar"218$}}
 \raise 2.0pt\hbox{$\mathchar"13E$}}}
\begin{document} 

\title{THE DYNAMICS OF GROUPS AND CLUSTERS OF GALAXIES \\
AND LINKS TO COSMOLOGY}

\author{
Gary A. MAMON \\ 
{\it Institut d'Astrophysique, 98 bis blvd Arago, Paris, F-75014 France}\\ 
also {\it DAEC, Observatoire de Meudon, Meudon F-92195, France}\\
gam@iap.fr}

\maketitle

\begin{abstract}
The links between the internal structure of galaxy groups and clusters
and cosmological parameters are reviewed here.
The mass density profiles of clusters, inferred from both optical analyses of
the 
galaxy surface number density profile coupled with internal kinematics, and
from weak gravitational lensing, appear cuspy with inner density 
profiles at least as steep as $r^{-1}$, as is expected from high-resolution 
cosmological
$N$-body simulations starting with scale-free or CDM initial conditions.
The high level of substructure both in the galaxy distribution and in the
diffuse hot gas seen in X-rays is consistent with dynamically young clusters
and a cosmological density parameter $\Omega_0 \gta 0.5$.
Despite the importance of cluster-cluster merging witnessed by these
substructures, the
spherical top-hat cosmology appears to give good indications of the
underlying physics of clusters and groups, including a fundamental
evolutionary track in a space of observational parameters for groups.
Dynamical processes operating in groups and clusters are reviewed.
Previrialization of groups and clusters before their full collapse 
is not expected on theoretical grounds, and the ensemble of
positions of compact groups,
relative to the fundamental track is a
potentially useful constraint on previrialization.

\end{abstract}

\parindent=0.25in
 
\section{Introduction}

This review focuses on the links between the richest hierarchies, {\sl
groups 
and clusters\/}, and global cosmological properties of the Universe,
principally the
density parameter, $\Omega_0$, and
the primordial density fluctuation spectrum, $P(k)$.
We begin with the observational facts, derive constraints on cosmological
parameters, and focus in the end on dynamical and cosmological evolution of
structures.
The review is written for non-experts, so the specialist should be patient
with some very obvious definitions given here.
We will not touch upon issues of the long term
evolution of galaxies in groups/clusters or the baryon content of these
systems. 

\section{Observational Facts}
\subsection{Small-scale structures of the Universe}
The attractive nature of gravity pushes matter to concentrate in overdense
regions, up to large scales, because of gravity's long range.
To first order, galaxies are a tracer of the total mass content of the
Universe (departures of which are called {\sl bias\/}).
And so, one expects and indeed sees that galaxies tend to agglomerate in
a hierarchy of structures as shown in Table \ref{hierarchy}.

\begin{table}[hbt]
\caption{Hierarchy of galaxy systems}
\begin{center}
\begin{tabular}{|l c c c c c c|}
\hline
System & $N_{\rm gal}$ & $m_{\rm faint}$ & Isolation & Scale &  
$\langle M/L \rangle$ & $P_{\rm gal}$ \\
\hline
Clusters & 30--300 & $m_3+2$ & None & $1.5 \, h^{-1} \, \rm Mpc$ & $300\,h$ &
10\% \\ 
Loose Groups & 3--30 & $m_1+3$ & None & $1.5 \, h^{-1} \, \rm Mpc$ & $150\,h$
& 50\%\\ 
Compact Groups & 4--8 & $m_1+3$ & $3 \,r$ & $<0.2 \, h^{-1} \, \rm
Mpc$ & $50\,h$ & 
0.1\%\\ 
Binaries & 2 &  $m_1+3$ & $5\,r_{12}$ & $<0.2 \, h^{-1} \, \rm Mpc$
& $50\,h$ & 10\% \\ 
Isolated & 1 & & & & & 30\% \\
\hline
\end{tabular}
\end{center}
\noindent Notes: $N_{\rm gal}$ is the number of galaxies per system within
the given range of magnitudes ($m = -2.5 \log_{10} flux + \rm cst$).
$M/L$ is the mass-to-light ratio, in solar units, assuming dynamical
equilibrium,  
$P_{\rm gal}$ is the fraction of galaxies in a given type of system, and
$h = H_0/(100 \,\rm km \,s^{-1} \, Mpc^{-1})$ is the dimensionless Hubble
constant, measuring the current rate of expansion of the Universe.
The criteria for clusters \cite{a58}, compact groups \cite{h82}, and
binaries \cite{t76}, have been slightly modified here.
\label{hierarchy}
\end{table}
Originally, galaxy systems have been cataloged through eye-ball searches of
galaxy positions projected on the sky.
Spectroscopy of galaxies provide redshifts, measuring galaxy radial
velocities (through the Doppler effect), which are the sum of the Hubble
expansion velocity (proportional 
to distance through the Hubble law $v = H_0 D$) and the {\sl peculiar\/}
velocity relative to this Hubble flow.
Except for nearby galaxies, the Hubble flow term dominates, so that redshift
is a first-order distance estimator.
Recent catalogs of galaxy systems are defined in 3D (projected positions
and depth inferred from the redshift) in automated fashion with well defined
criteria.

One should understand that galaxy systems probably span a continuous range of
hierarchies, and therefore not put too much weight in the special categories
defined here, whose properties depend strongly on the adopted selection
criteria.
{}From Table \ref{hierarchy}, most galaxies lie within fairly rich
environments, and we shall see in \S\ \ref{tophat} that they are
dense enough to be, using cosmological terms, in the quasi-linear to
non-linear regimes of the growth of 
the primordial density fluctuations by gravitational instability.
Their properties are thus function of both the underlying cosmological
parameters and the dynamical processes operating within them.

\subsection{Density profiles}
\label{obs-density}

Surprisingly little is known on the galaxy number or total mass
density profiles in groups and clusters?
The galaxy number density profiles of groups and clusters are poorly known
because of 
\begin{itemize}
\item small number statistics (most clusters have at best a few hundred
galaxies with radial velocities confirming their membership)
\item their unknown center
\item contamination from foreground and background interlopers
\end{itemize}

For example, in clusters, 
model profiles with homogeneous cores fit the
data \cite{b75}, but so do models with 
cuspy
cores \cite{y74}, and the two are hard to distinguish 
given the limits on the data \cite{y74}.
Homogeneous cores are the best fits when one superposes the information from
different clusters, rescaling to the same size \cite{bt86}, but when
one uses the X-ray barycenter or the position of the giant {\sl cD\/}
galaxy as the cluster center instead of the center of the galaxy positions, 
the clusters  unequivocally show cuspy cores \cite{bt86}.
So homogeneous cores were the result of poor choices for the cluster
centers.
The cuspiness of clusters has been recently confirmed \cite{mg94}
in an analysis of 1500 suspected members of a single nearby rich
cluster (Coma), for which
the number space-density profile  is $\nu \sim r^{-1}$.
The presence of possible interlopers in the sample can only wash out the
central density cusp, so that, in reality, the slope of the 3D number density
profile may be even steeper.

If the situation in galaxy clusters is difficult, it should be impossible to
conclude anything on groups of galaxies.
Nevertheless, by averaging over groups in the NBG \cite{t87}
catalog, and rescaling the groups to some unit size, it has been
noted \cite{wm89} that these were closer to
homogeneous than to $\nu \sim r^{-2}$.
Averaging over the best defined compact group sample \cite{h82}, again with
rescaling, 
compact groups catalog were found to be significantly centrally
concentrated \cite{hnhm84}, although the slope of the underlying profile is
not given.
A reanalysis \cite{mdgms96} of that compact group sample, after removal
of groups with interlopers in redshift space \cite{hmhp92}, and
starting 
with the distribution of absolute pair separations, shows that
the distribution 
of galaxy separations within compact groups is consistent with a unique
absolute density profile, falling off as $\nu \sim r^{-2.4}$ in the envelope
and with a small homogeneous core of size $18\,\, h^{-1} \, \rm kpc$, that is
half the median pair separation within compact groups.
It is not yet clear that this result is caused by high central
concentration or by the presence of tight binaries within less concentrated
density profiles (as seems to be the case in loose groups \cite{wm89}).

What about the total mass density profiles?
One can resort to either optical data or X-ray data.
In both cases, one writes the equation of hydrostatic equilibrium (or Jeans
equation), which in its general spherically symmetric form is
\begin{equation}
{dP\over dr}= - \rho {d\Phi\over dr} = \rho {GM_{\rm tot}(r) \over r^2} \ ,
\label{hydro_simple}
\end{equation}
where $\rho$ is the density (in mass or number) of whatever tracer we use to
measure the pressure $P$.
It is then easy to obtain the mass density profile
\begin{equation}
\rho = {1\over 4\pi r^2} {dM_{\rm tot} \over dr} \ .
\label{density}
\end{equation}

Now, from equation (\ref{hydro_simple}),
the Jeans equation for the `gas' of galaxies in a spherical cluster is 
\begin{equation}
\nu{d\sigma_r^2\over dr} + 2\beta_{\rm anis} {\nu\sigma_r^2\over r} = 
-\nu {GM_{\rm tot} (r) \over r^2} \ ,
\label{hydro_opt}
\end{equation}
where $\nu$ is the space number density of galaxies in the cluster,
$\sigma_r$ is the radial velocity dispersion (standard deviation of the
velocity distribution) of the cluster, and $\beta_{\rm anis} = 1 -
\sigma_t^2/\sigma_r^2$ is the velocity anisotropy, given that $\sigma_t$ is
one tangential component of the cluster velocity dispersion.

The problem with the optical analysis is that the variation of
anisotropy with radius is unknown, even if there are good reasons to believe
that, in the central regions of clusters, the system should be isotropic,
since the two-body relaxation time of galaxy-galaxy encounters is short (see
\S\ \ref{timescales} below).
The anisotropy of the outer envelope can be either radial, if the envelope
evolution is dominated by near-radial infall of galaxies, or tangential if
the cluster is experiencing an off-center collision with a smaller one.
Note that the reversed problem of fixing the mass density profile and
computing the anisotropy profile is solvable in quadrature \cite{bm82}.

The advantage of the X-ray method is that there is no pressure anisotropy
term to include in the Jeans equation.
Using the equation of state of an ideal hot X-ray emitting gas, the total
mass profile is \cite{flg80}
\begin{equation}
M_{\rm tot} (r) = -{ k T r \over G \mu m_p } \left (
{d\ln n\over d\ln r}
+
{d\ln T\over d\ln r}
\right ) \ ,
\label{mtot_X}
\end{equation}
where $n$ and $T$ are the local gas number density and temperature, while
$\mu m_p$ is the mean particle mass in the hot plasma.
Since both the galaxy system and the gas reacts to the same overall
potential, and when both systems are isothermal, and
when the galaxy system is also isotropic, one finds \cite{c74}
(eqs. [\ref{hydro_opt}] and [\ref{mtot_X}]) $n_{\rm gas}
\sim n_{\rm gal}^\beta$, where $\beta = \sigma_v^2 / (kT/\mu m_p)$ is the
ratio of kinetic energies of the galaxies and gas.
Now, using equations (\ref{density}) and (\ref{mtot_X}), one can compare the
total mass density
profile to the gas density profile, where the emissivity of the gas locally
scales as the square of its density.
It turns out that the dark matter is significantly more centrally peaked than
the diffuse X-ray emitting gas \cite{h85,h89,gdll92,dglls94}. 
However, the dark matter density profile is similar to the galaxy number
density profile \cite{h89,dglls94}.

The optical analysis of the Coma cluster, assuming isotropy of the velocity
distribution, 
yields \cite{mg94} a total mass 
density $\rho \sim r^{-3}$ steeper than the galaxy number density given
above.
Putting in
radial anisotropy at large radii produces a broken total mass
density profile varying as $r^{-3}$ in the center but only $r^{-1}$ in the
envelope \cite{mg94}. It seems difficult to have the inner mass density
profile fall as slowly as the galaxy number density profile with an
anisotropy profile consistent with dynamical and cosmological principles.
If the Coma cluster is indeed regular, the X-rays may be telling us something
about the radial variation of the velocity anisotropy.

Gravitational optics is rapidly becoming a promising alternative method for
deriving projected mass distributions.
Indeed, 
the amplification due to an elliptical gravitational lens is \cite{bk75}
\begin{equation}
A = {1\over (1-\kappa)^2 - \gamma^2} \ ,
\label{eq:amp}
\end{equation}
where the matter (or Ricci) term $\kappa$ can be expressed in terms of a
critical surface mass density $\Sigma_c$:
\begin{eqnarray}
\kappa &=& {\Sigma_{\rm mass}(R)\over\Sigma_c} \ ,
\label{eq:kappa}\\
\Sigma_c &=& {c^2\over 4\pi G} {1\over D_l (1 - D_l/D_s)} 
\ , \nonumber
\label{eq:sigmac} 
\end{eqnarray}
where $D_{l}$ and $D_{s}$ are the distances to
the lens and to the source, respectively,
and where the shear (or Weyl) term is
\begin{equation}
\gamma = \kappa - \frac{2}{R^2 \Sigma_c}\int_0^R \Sigma(x) x dx \ .
\label{eq:gamma}
\end{equation}
Equations (\ref{eq:amp}) and (\ref{eq:kappa}) indicate that
gravitational amplification of background galaxies by the potential of a
galaxy group  cluster is directly related to the surface mass
density.

When the surface mass density of the cluster is close to $\Sigma_c$, 
background galaxies are lensed into a giant arc, arclet, or are just
slightly elongated, in a sequence of decreasing alignment of the background 
galaxy with the caustics in the plane of the foreground
cluster \cite{gn88}.
While giant arcs have been observed \cite{sfmp87,lp86}, there only comes
one or two per cluster and thus cannot constrain the cluster density profile.
On the other hand, the elongations of lensed background galaxies, caused by
the shear term (eq. [\ref{eq:gamma}]), allow one to {\it map\/} the surface
density of a cluster, by measuring and inverting the shear field \cite{ks93}.
With such methods, one finds cuspy surface density profiles at least as steep
as $\Sigma
\sim r^{-1}$  \cite{tf95,skbfwnb95}.
Surprisingly, there often seems to be more mass at a given radius than inferred
from the X-rays (eq. {\ref{mtot_X}]) by a factor 2 to
3 \cite{lm94,mb95}, but the 
discrepancy disappears if one allows for small-scale fluctuations in the
radial variation of the gas temperature \cite{skbfwnb95}.

\subsection{Substructure}
\label{obs-substruct}

The na\"{\i}ve view that clusters of galaxies are regular structures with
little substructure can be taken as justification for the dynamical analyses of
their underlying mass content.
Recent studies point to the opposite picture of very irregular
clusters.
These are based upon analyses of the projected distribution of galaxies in
clusters without \cite{ans64,b77,gb82} or with \cite{bgbh83}
the radial velocities of the member galaxies, as well as on
X-ray \cite{jf92} analyses.
In particular, at least 30\% of clusters show significant substructure from
optical 3D analyses \cite{ds88}, including the once canonical regular Coma
cluster
 \cite{b84}.
Similarly, the X-ray analyses indicate that at least 22\% of clusters
show bimodal substructure \cite{jf92}, and optical analyses indicate that
roughly half \cite{sgs93} or as much as 90\% \cite{ebggmmm94}
of the clusters that are regular in the X-rays 
show small-scale substructure (the latter analysis is based upon wavelets).

In loose groups, again the small-number statistics make it very difficult to
assess statistically significant substructure.
In a system of say 8 galaxies, small-scale substructure would show up as
galaxy pairs (subgroups of 
triplets or quartets of galaxies would show up as separate groups in the
catalogs).
For the NBG \cite{t87} group catalog, 1.5 pairs are found on average per group
 \cite{wm89}, although a fraction three times lower has been
claimed \cite{hr88} for groups in an unpublished catalog.

\subsection{Fundamental plane}
\label{fp}

Three parameters are most easily measured in astronomical systems:
angular size, velocity dispersion (or temperature), and flux.
In a subsample of systems at known distances, angular size provides the
physical size, and flux provides the (intrinsic) luminosity ({\it i.e.\/,}
power).
Thus, astronomical systems are well characterized by the parameters, $R$
(size), $V$ (velocity dispersion), and $L$ (luminosity).
Whether these systems are star clusters, galaxies, or galaxy clusters, people
have attempted to find a plane in $R,V,L$ space in which the objects of a
sample all fit.
Such {\sl fundamental planes\/} have indeed been found for galaxy
clusters \cite{smcb92} with the relation $L \sim R^{0.9} \,V^{1.3}$, not very
different from that of elliptical galaxies (with no disks or spiral
structure) or globular star clusters.

Fundamental planes are useful for two reasons:
1) They teach us about how the systems form and evolve.
2) They serve as distance indicators, independently of the Hubble law ($V_r =
H_0 D$, where $V_r = cz$ is the radial velocity, while $c$ is the velocity of
light, and $z$ the redshift).
This is important, because the Hubble law is only uniform on average on very
large scales, but on smaller scales is perturbed by local overdensities of
matter (where the expansion rate is lower than the average one,
and becomes even negative when matter falls into a galaxy, group, or cluster)
and underdensities (where the expansion rate is larger).
Therefore, outliers from the fundamental plane are usually systems for
which the 
distance was poorly estimated, and the correct distance is obtained by
forcing the system to lie on the plane.
In other words, if $L \simeq {\rm Cst} R^\alpha \,V^\beta$ (no relation with
the previous $\alpha$s and $\beta$s), then the correct
distance to an outlier is
$$
D \simeq \left ({ {\rm Cst} \,\theta^\alpha V^\beta \over 4 \pi f} \right
)^{1/(2-\alpha)}
\ ,
$$
where the angular size $\theta = R/D$, and the measured flux $f = L/(4\pi
D^2)$.

\subsection{Internal kinematics}

The motions of galaxies in clusters probe the potentials of these systems,
when these are near dynamical equilibrium.
For example, Jeans' equation 
(\ref{hydro_opt}) specifies the radial variation of the
second moment of the velocity distribution, the squared velocity dispersion.
The measured line-of-sight velocity dispersion is the
emissivity-weighted average of the velocity dispersion, where the geometry
becomes a little complicated in the presence of velocity anisotropy \cite{bm82}.
In any event, the radial variation of the line-of-sight velocity dispersion
provides the knowledge of the velocity anisotropy, assuming that one
knows the potential \cite{bm82}, or of the potential if one knows the
anisotropy.

The early realization that the velocity dispersion decreases towards the
center of
the Coma cluster \cite{kg82} (as well as outwards in the envelope)
is interpreted as either a
cuspy total mass density profile with a core that has isotropic
velocities \cite{b82} or a homogeneous core with 
tangential (nearly circular) velocity dispersions \cite{hk95}.
A recent systematic study of a large number of clusters \cite{hk95}
reveals that roughly 15\% of clusters have such inverted velocity dispersion
profiles, while roughly 40\% have velocity dispersions decreasing outwards
everywhere, the remaining profiles being either flat or unclassifiable.
The clusters with cuspy velocity dispersion profiles indicate cuspy steep
density
profiles ($\rho \sim R^{-\gamma}$, where $\gamma = 3.5-4$) \cite{hk95}, 
with at best very small ($<40 \, h^{-1} \, \rm kpc$) homogeneous
cores \cite{hk95}.


\section{Cosmological Constraints}

\subsection{The spherical top-hat approximation}
\label{tophat}

Consider a constant finite spherical overdensity in a homogeneous 
Universe.
The evolution of this {\sl top-hat\/} perturbation is governed by the simple
equation 
\begin{equation}
\ddot R = -GM/R^2 \ .
\label{eqmom}
\end{equation}
It will 
expand with the Hubble expansion then reach maximum expansion at the
{\sl turnaround\/} time and collapse at twice the turnaround time.
Subsequently, the matter may reach dynamical equilibrium, and become 
{\sl virialized\/} as it satisfies
the virial theorem $2\langle T \rangle + \langle W \rangle = 0$, where $T$
and $W$ are its kinetic and potential energies, respectively.

It is easy to show (without cosmological arguments, simply integrating
eq. [\ref{eqmom}]) that the turnaround time
can be expressed in terms of the density at turnaround
\begin{equation}
\tau_{\rm ta} = \left (32 G \rho_{\rm ta} \over 3\,\pi\right)^{-1/2} \ ,
\label{eq:tta}
\end{equation}
where $\rho_{\rm ta}$ is the density at turnaround.
Once the top-hat collapses, the surrounding regions will collapse onto it in
a gradual fashion, each reaching maximum expansion following equation
(\ref{eq:tta}) where $\rho_{\rm ta}$ is now the mean density within the
turnaround region at the epoch of turnaround \cite{gg72}.
This process of so-called {\sl secondary infall\/} has been studied in
detail \cite{gg72}, and for $\Omega = 1$, one
obtains simple scaling laws for the matter within the {\sl turnaround\/}
radius, beyond which matter is expanding:
\begin{equation}
M_{\rm ta} \sim t^{2/3} \qquad R_{\rm ta} \sim t^{8/9} \qquad \bar \rho_{\rm
ta} \sim t^{-2} \ .
\label{ta}
\end{equation}

Reversing equation (\ref{eq:tta}) one can then obtain the present radius of
turnaround of cosmological objects in the top-hat approximation, and these
are given in Table \ref{tb:rta} from 
\begin{equation}
R_{\rm ta} = \left (8 G M_{\rm ta} t_0^2 \over \pi^2\right)^{1/3} \ ,
\label{eq:rta}
\end{equation}
independent of $\Omega_0$ \cite{gg72}.
Here, $t_0$ is the age of the Universe
\begin{table}[hbt]
\caption{Turnaround radii}
\begin{center}
\begin{tabular}{|l c c|}
\hline
System & $M_{\rm ta} (M_\odot)$& $R_{\rm ta}$ (Mpc)\\
\hline
Small galaxy	& $3\times 10^{11}$	&	0.4	\\
Large galaxy	& $3\times 10^{12}$	&	0.8	\\
Small group	& $3\times 10^{13}$	&	1.7	\\
Small cluster	& $3\times 10^{14}$	&	3.6	\\
Large cluster	& $3\times 10^{15}$	&	7.8	\\
\hline
\end{tabular}
\end{center}
\label{tb:rta}
\end{table}

Within the turnaround radius, matter is infalling up to the region of shell
crossing inside which matter is well mixed (see \cite{m92} for a graphical
example). The radius of mixing is approximately the radius of second
turnaround or {\sl rebound radius\/}, which is roughly one-third of the
turnaround radius \cite{b85}.
For an $\Omega = 1$ universe, the density is critical at $\rho_c =
(6 \pi G t^2)^{-1}$, and the mean overdensity within the turnaround radius is
(using eq. [\ref{eq:rta}]) $9\pi^2/16 = 5.55$.
If matter virializes at collapse and settles, from the virial theorem, at
half its turnaround radius at epoch $2\,\tau_{\rm ta}$, its density will be 8
times larger than it was at turnaround, while the density of the
Universe will decreased by a factor (eq. [\ref{ta}]) 4 (for $\Omega = 1$), so
that the overdensity at virialization is $18\,\pi^2 = 178$.
Loose groups of galaxies are defined with density contrasts of 
20 \cite{gh83} to 80 \cite{rgh89}, and so must still be feeling the collapse.
Combining the virial theorem with this cosmological minimum density yields
a small maximum size of $\simeq 200 \, h^{-1} \, \rm kpc$ and minimum
velocity dispersion of $\simeq 250 \, \rm km \, s^{-1}$ for groups \cite{m94}.

Recent cosmological collisionless $N$-body simulations \cite{cl95} with
power-law primordial 
density fluctuation spectra $P(k) \sim k^n$ indicate that the spherical
top-hat approximation is very good, in that the radius 
where the density is 178 times the mean density delimits the infalling
region from the quasi-static region.
Nevertheless, the top-hat approximation cannot represent all the cosmological
aspects of clusters and groups, because the Universe is filled with
structures and substructures (\S\ \ref{obs-substruct}).

\subsection{Density profiles}

\label{th-density}

What should one expect for the density profiles of clusters?
In simple terms, the final density profile of an object will be set in part
by whether the object collapses at once or in a slower smoother fashion.
The details are set by the primordial density fluctuation spectrum, {\it
i.e.\/,} 
the variation of the rms overdensity fluctuations $\sigma_\delta$ with
(comoving) scale $R$.

In the top hat approximation (\S\ \ref{tophat}),
it is easy to derive density profiles using the relations of 
equation (\ref{ta}), yielding $\rho \sim r^{-9/4}$ and $\sigma_v \sim
r^{-1/8}$ \cite{g75}.
The $-9/4$ slope of the density profile has been verified with more detailed
semi-analytical calculations \cite{fg84,b85} and with $N$-body
simulations \cite{mabp95}.

Consider now a spherical top-hat perturbation in an open ($\Omega < 1$)
Universe. Because the Universe is less dense than critical, there will be a
lack of matter falling onto the perturbation at late times, so that the
envelope of the evolving system should be steeper than $\rho \sim
r^{-9/4}$ \cite{g77},
as checked with $N$-body simulations \cite{zqs88,cer94}.
However, the regions that are accreted at early times came from a Universe
whose density was closer to critical, thus resembling more an $\Omega = 1$
Universe.

Now reality is not as simple as spherical top-hat perturbations in an otherwise
homogeneous Universe.
First, perturbations may not be spherical, which then produces shallower
slopes \cite{fg84}.
Moreover, fluctuations in the primordial density field occur everywhere
on a variety of scales.
One can then show analytically
 \cite{hs85} that the density profiles have slopes of 
${\rm Min} [-2, -3(3+n)/(4+n)]$, given a scale-free primordial density
fluctuation 
spectrum $P(k) \sim k^n$, and again, this has been checked with $N$-body
simulations \cite{efwd88,zqs88,cer94}.
$N$-body simulations with CDM spectra (which have spectral index $n$
increasing with scale from $-3$ to 1, and is roughly $-2$ on the scale of
galaxies and $-1$ on the scale of clusters) show
a varying a slope from $-1$ inside to $-4$ \cite{dc91} or $-3$ \cite{nfw95}
outside
(the latter study 
includes dissipative gas).
Interestingly, high-resolution scale-free simulations \cite{cl95}
reproduce the density profile with slope varying from $-1$ to $-3$ found in
low resolution CDM simulations with gas \cite{nfw95}.
All the
high resolution simulations with CDM or scale-free spectra show cuspy cores,
with a core radius smaller than the numerical resolution, which can be as
small as a few kpc \cite{dc91}.

\subsection{Substructure}

In a low density Universe, structure formation tends to freeze at redshift $z
= 1/\Omega_0$ \cite{gr75}, because there is simply not enough matter to keep
accreting onto already formed structures.
In contrast, a critical density Universe sees accretion at all times.
Therefore, the level of substructure in galaxy clusters seems to be an
excellent way to constrain the density parameter $\Omega_0$.

The approach is to predict the fraction of galaxy clusters with given level of
substructure, as a function of $\Omega$ and compare with the observations (\S\
\ref{obs-substruct}).
One way is the analytical computation of the distribution of collapse times of
structures of given mass \cite{rlt92}.
Alternatively, one builds Monte-Carlo realizations of the history of structures
and their mergers, based upon the Press-Schechter \cite{ps74} mass function 
of structures
at given epoch \cite{lc93,kw93} (this is, in essence, a different extreme
representation of the Universe than the top-hat model, which is limited to
the accretion of numerous small objects onto groups and clusters).
All studies yield $\Omega_0 \gta 0.5$, although
the precise lower limit to the density parameter is still subject to debate.
These predictions are insensitive to the primordial density fluctuation
spectrum \cite{lc93} and to the cosmological constant \cite{rlt92}.

For example, bimodal substructure is expected for a critical density universe
in 15\% \cite{kw93}, 24\% \cite{rlt92}, and 28\% \cite{lc93}, depending on how
long the bimodal 
substructure can survive within the cluster (assumed at 0.08 times
the present age of the Universe, for the first two studies and $0.1
\,t_0$, respectively), 
compared to $>22\%$ observed (\S\ \ref{obs-substruct}).
For $\Omega_0 = 0.2$ the corresponding predicted fractions are 4 to 6 times
lower, and clearly discrepant with the observations.

Half of all clusters are predicted to harbor 10\% level small-scale
substructure \cite{kw93} if $\Omega_0 = 1$, which compares well with the
smaller of the two observed estimates \cite{sgs93}.
If instead 90\% of all clusters harbor small-scale
substructure \cite{ebggmmm94}, one would need either a higher $\Omega_0$ or a
survival time longer than $0.08\,t_0$.
A recent study \cite{gms94} does provide a 50\% longer survival time for
small-scale substructure caused by accreted small clusters, but slightly
less, when the small-scale substructure if caused by the dense cores of two
similar mass clusters detaching from their parents when these two merge
together \cite{gss93}.

$N$-body simulations with gas \cite{emfg93,mefg95} confirm the
semi-analytical 
predictions given above and illustrate very well how the
gaseous distribution in clusters in a low-$\Omega$ Universe is considerably
smoother on small-scales than observed in the X rays.

\subsection{Previrialization}
\label{previr}

Most analytical approaches to cosmology apply the top-hat approach, in which
the collapse is radial.
In principle, non-radial motions can arise as angular momentum can be pumped
into the system by external tidal torques or by gravitational interactions on
small scales within the system.
The existence of significant non-radial motions is known as {\sl
Previrialization\/} \cite{dp77}.
%
%
How important is this effect?
The first numerical study \cite{p90} argued for a very strong effect from the
outer environment, so that
the critical density for collapse, expressed in units of present day
linearly evolved primordial density, is found to be of order 5, instead of
the canonical value of 1.69 for $\Omega_0 = 1$ \cite{p80} or 
1.61 for $\Omega_0 = 0.1$ \cite{lc93}.
A more realistic set of cosmological $N$-body simulations \cite{ec92}
with $P(k) \sim k^{-1}$, showed that power on small scales within clusters
does not alter its collapse. Moreover, implicit in this study is that 
external tidal torques bring only
low angular momentum ($J = 0.07\,J_{\rm circ}$) just within
the turnaround radius of clusters.
A very recent analytical study \cite{ljbh96}, using perturbative expansions
in the quasi-linear regime, suggests that for $n \simeq
-1$, the collapse is as one expects from linear theory, hence with
insignificant non-radial motions.
The cold dark matter spectrum, still considered to be a close approximation
to the true power spectrum \cite{wssd95}, has a slope close to 
$-1$ for the 
infalling regions of groups and clusters, hence previrialization should not
be important in modeling these systems.

\section{Dynamical Evolution}
\label{timescales}

There are quite a few dynamical processes competing in the evolution of
groups and clusters of galaxies. 
What is written below is largely taken from a previous review \cite{m93}.

The simplest timescale in a self-gravitating system is the circular orbital
time: 
\begin{equation}
\tau_{\rm orb} = {2\pi R\over V_{\rm circ} (R)} = \left [{3\pi \over G \bar
\rho (R)} \right ]^{1/2} \ ,
\label{eq:orb}
\end{equation}
since $V_{\rm circ}^2 = G M(R)/R$.

\subsection{Relaxation}

\label{sec-relaxation}

As a test particle undergoes scattering collisions within a sea of field
particles, it will progressively forget its initial conditions.
This {\it two-body relaxation\/} time can be defined in at least three ways:
$$
\tau_{\rm 2-rel} \equiv 
\left\langle {1 \over v^2}{dV^2 \over dt}\right\rangle^{-1}
\hbox {   or   }
\left\langle {1 \over E}{dE \over dt}\right\rangle^{-1}
\hbox {   or   }
\left\langle {d\sin^2 \Delta\alpha\over dt}\right\rangle^{-1}
\ ,
$$
where $\Delta \alpha$ is the deflection angle in an encounter.
This can be written as \cite{c42}
$$
\tau_{\rm 2-rel} = {v^3 \over G^2 m_f^2 n f(v/\sigma_v) \ln \Lambda} \ ,
$$
where $v$ is the velocity of the test particle, 
$m_f$, $n$, and $\sigma_v$ are the mass, number density, and 1D velocity
dispersion of the field particles, respectively,
$f(x)$ is the fraction of particles traveling faster than $x\,\sigma_v$, 
and $\ln \Lambda$, is called the {\sl Coulomb
logarithm\/}, with typical values of 2 to 10, where $\Lambda$ is the ratio of
maximum to minimum impact 
parameter.
%
For a system of galaxies and dark matter
particles, one finds that the galaxies relax by galaxy-galaxy collisions,
but not by collisions with
individual dark matter particles (whose masses are too low).
Similarly, the dark matter particles relax mainly by collisions with
individual galaxies.

A {\sl collective relaxation\/} time has been derived,
not by summing up the encounters but by computing the collective
response of the system \cite{gs86}
$$
\tau_{N\rm -rel} = {\rm Cst} \, {v \over G m_f n^{2/3}} \ .
$$
This collective relaxation turns out to be somewhat more efficient than
two-body 
relaxation in clusters and loose groups but not in dense groups.
In general, only the cores of rich clusters are relaxed.

Particles that evolve in a rapidly time varying potential
can rapidly forget their initial conditions \cite{lb67}
This {\sl violent relaxation\/} occurs in a timescale
$$
\tau_{\rm v-rel} \sim \tau_{\rm ff} \sim \tau_{\rm dyn} 
\quad\hbox{   when   }\quad
|{\partial \phi \over \partial t}| > |{\bf v} \cdot \nabla \phi| \ ,
$$
where $\tau_{\rm ff}$ is the free-fall time, and $\phi$ is the global
potential.
This applies for example to collapsing systems, as is often the case in
cosmology.
For example, elliptical galaxies are thought to form by dissipationless
collapse or by mergers of disk galaxies \cite{t77} \cite{m92}, and since both
phenomena perturb the potential in a rapid violent manner, the cores of
elliptical galaxies appear relaxed
although their 2-body (and collective) relaxation times are much longer than
the age of the Universe.

Energy exchanges during encounters lead to {\sl energy equipartition\/}, and
the more 
massive objects tend to move slowly and fall to the center of their systems,
leading to {\sl mass segregation\/}.
The timescale is similar to that of 2-body relaxation.

Occasionally, collisions will pump sufficient energy into a particle that
it's velocity will be larger than the escape velocity of its system, and,
barring subsequent encounters that may reduce its kinetic energy, the
particle will escape. If during one relaxation time the distribution of
particle velocities reaches a Maxwellian, the fraction of unbound particles
is roughly 1\% so that 
the timescale for {\sl evaporation\/} is roughly 100 times the relaxation
time \cite{a38}.

\subsection{Dynamical friction, orbital decay and circularization}

Field particles are deflected by the mass of the test particle.
Hence,
in the frame of the
test particle, the field particle density is higher behind the test particle
than in front of it. This leads to a drag force known as
{\sl dynamical friction\/} \cite{c43}, which plays a major role in group and
cluster dynamics. For an infinite homogeneous medium, the timescale for
dynamical friction is \cite{c43}
\begin{equation}
\tau_{\rm df} \equiv 
\left ({1 \over v_\parallel}{d v_\parallel \over dt}\right )^{-1}
= {v^3 \over 2 \pi G^2 (m+m_f) \rho f(v/\sigma_v) \ln (1+\Lambda^2) }\ ,
\label{eq:df}
\end{equation}
where $\rho$ is the local mass density of field particles, and where $f$ and
$\Lambda$ are defined in \S\ \ref{sec-relaxation}.

Perhaps more physical is the timescale for {\sl orbital decay\/} defined as
\begin{equation}
\tau_{\rm od} \equiv \left ({1 \over R}{dR\over dt}\right)^{-1}
= \left({RdE/dR\over mv^2}\right) \tau_{\rm df}
= {3\over2}\left ({\rho\over \bar\rho} + {1\over 3}\right )\tau_{\rm df} \ .
\label{eq:od}
\end{equation}

Dynamical friction and orbital decay can lead to misleading answers:
\begin{itemize}
\item No orbital decay is predicted in zero density environments,
whereas a satellite galaxy sitting just outside its parent galaxy will see its
orbit decay, because of resonances with its parent 
 \cite{lt83}
\item
Although orbital decay should be slowed by tidal effects
that reduce the test particle's mass, the contrary may occur with a satellite
galaxy circling its parent, as the tides from the latter remove stars from the
former, and these carry off energy and angular momentum, thus accelerating the
orbital decay
 \cite{pc92}
\item For a particle radially falling into a medium with
a outwards decreasing density profile, the dynamical friction time computed
from equation (\ref{eq:df}) is longer than the time on which the particle
sees an increasing density.
A system, dense enough to survive the tidal forces from the primary into which
it falls, will nearly stick to the core of the primary after its second
passage, if its mass is more 
than 10\% of the primary's mass \cite{cgm95}.
\end{itemize}

For circular orbits, combining equations (\ref{eq:orb}), (\ref{eq:df}), and
(\ref{eq:od}),
one obtains 
\begin{equation}
{\tau_{\rm od} \over \tau_{\rm orb}} = {\left (1 + {1\over3} \bar\rho/\rho
\right ) M/m \over 2\pi f(1) \ln (1+\Lambda^2) (1+m/M)^{1/2}}
\simeq {2 \over \pi }{M/m \over (1+m/M)^{1/2} \ln \left[1+(M/m)^{2/3}\right]}
\ ,
\label{eq:odoverorb}
\end{equation}
where $f \simeq 1/2$, 
$p_{\rm min} \sim r$, 
$p_{\rm max} \sim R$,
primary and secondary have the same mean density ($[M/m] \sim [R/r]^3$), 
and for singular isothermal density profiles ($\rho
\sim r^{-2}$), $\bar \rho = 3\rho$.
The term $(1+m/M)^{1/2}$ is a correction for non-negligible secondary to
equations (\ref{eq:orb}) and (\ref{eq:df}).
Table \ref{tb:od} below gives the ratios from equation (\ref{eq:odoverorb})
for typical astrophysical ratios.
\begin{table}[hbt]
\caption{Number of circular orbits for orbital decay}
\begin{center}
\begin{tabular}{|l r r|}
\hline
Secondary	&	$M/m$		&$\tau_{\rm orb}/\tau_{\rm od}$\\
\hline
Galaxy		&	$10\,000$	&	1000\\ 
Group		&	$100$	&	20\\
Small cluster	&	$10$	&	3.5\\
Moderate cluster&	$3$		&	1.5\\
\hline
\end{tabular}
\label{tb:od}
\end{center}
\noindent Notes: Examples of secondaries are
for a rich cluster primary, but should be considered as academic, since the
secondaries are not expected to acquire sufficient angular momentum as to
orbit in near circular orbits as
they enter the cluster (\S \ref{previr}).
\end{table}

{}From table \ref{tb:od}, dynamical friction and orbital decay are not
important, in terms of number of near-circular orbits, unless the mass ratio
of secondary to primary is $\gta 0.1$, {\it i.e.\/,} small clusters falling
into larger ones, or galaxies falling into small groups.
In absolute terms, the circular orbital time is $\tau \simeq 0.4 t_0$
for a 
secondary orbiting a $10^{15} M_\odot$ cluster at $1 \, h^{-1} \, \rm Mpc$
from its center.
Hence, the
timescale for orbital decay in rich clusters is greater than
the age of the Universe for systems with $m < 10^{14} M_\odot$, orbiting at
$1 \, h^{-1} \, \rm Mpc $.

Another outgrowth of dynamical friction is {\sl orbital circularization\/},
whose timescale can be defined as the rate a test particle acquires angular
momentum from interactions with other particles:
$$
\tau_{\rm oc} = \left ({1 \over J_{\rm circ}(E)} {dJ\over dt}\right)^{-1}
\ ,
$$
and is found \cite{m85}
to be shorter than the orbital decay time outside of
the core of a cluster.

\subsection{Tides}

Tidal forces act on particles in a system {\it relative\/} to the full system
itself. As such there are two types of tides acting on galaxies in groups and
clusters: those caused by close encounters with other galaxies and those caused
by variations in the gradient of the global group/cluster potential.
The first type of tides ({\sl collisional stripping\/}) has a timescale
$$
\tau_{\rm cs} \equiv \left ({1 \over m}{d m\over dt}\right )^{-1}
= \left\langle (\Delta m / m) n \langle \sigma v \rangle\right\rangle^{-1}
= {{\rm Cst} \over n r_g^2 v_g} \ ,
$$
where $\sigma$ is the collisional stripping cross-section, and the outer stars
are assumed to follow elongated orbits \cite{r75,dls80}.

Global potential tides depend strongly on the galaxy's orbit around the
cluster. If the galaxy is phase
locked in a nearly circular orbit around the cluster, it will feel a roughly
constant tidal 
shear, and its tidal radius will be obtained by equating the tidal shear at a
given radius in the galaxy with the gravitational pull that the full galaxy
exerts on a star at that radius, plus an inertial term:
\begin{equation}
\Delta \left ({GM(r)\over R^2}\right ) = -{Gm(r) \over r^2} + \Omega^2 r \ ,
\label{eq:tide}
\end{equation}
yielding for $r \ll R$
\begin{equation}
\bar \rho_g(r_t) = \bar\rho_{\rm cl}(R) \left [2 - 3 \,{\rho_{\rm cl} (R) \over
\bar \rho_{\rm cl} (R)} + {V_p^2 (R) \over V_{\rm circ}^2 (R_p)}\right]
\ ,
\label{eq:denscrit}
\end{equation}
{\it i.e.\/,} the galaxy is tidally truncated at a radius $r_t$
where its mean density
is of the order of the mean cluster density within the radius $R_p$ of closest
approach of the galaxy (where $V_p$ and $V_{\rm circ}$ are the pericentric and
circular velocities, respectively).

If the orbits are elongated, the instantaneous tide obtained from equation
(\ref{eq:tide}) is short lived and the galaxy experiences a {\sl tidal
shock\/} \cite{osc72}. 
Using the impulse approximation \cite{s58},
in which the perturber moves with a constant relative velocity $\bf V$,
one can show \cite{m92},
again for $r \ll R$, that
$$
\bar \rho_g (r_t) = {\rm Cst} \,\bar \rho_{\rm cl} (R_p) f(\epsilon) \ ,
$$
where $R_p$ is the pericentric of the galaxy's orbit, and $f(\epsilon)$ is a
function of order unity of the galaxy's orbital eccentricity.
This criterion is similar to that for circular orbits, but the constants are
higher, because at given pericenter, a galaxy in a circular orbit must feel a
more effective tide, since it is
long-lived \cite{m87}.
Numerical simulations \cite{ar88}
confirm this result although
other simulations \cite{mw87}
suggest that the  tide is most
efficient for some intermediate elongation at given pericenter, when this is
within the nearly homogeneous region of the cluster.
Note that the timescales for global potential tides are basically the
orbital timescales divided by the typical mass-loss per passage through the
cluster core.

The effectiveness of a tide is related to the maximum strength of the tide
times the duration of this maximum tide.
So, from equation (\ref{eq:denscrit}) one gets
\begin{eqnarray*}
\Delta v \sim F_{\rm tid} \Delta t &&\sim \bar \rho_g \Delta t
\sim { 2 - 3 \rho_{\rm cl} / \bar \rho_{\rm cl}  + V_p^2 / V_{\rm circ}^2
\over V_p / V_{\rm circ} } \\
&&\sim 3\left(1-{\rho_{\rm cl} \over \bar \rho_{\rm cl} }\right) 
- \left (1-3{\rho_{\rm cl} \over \bar \rho_{\rm cl} }\right) 
\left ({V_p \over V_{\rm circ}} -1\right) \quad \hbox { for } V_p \gta
V_{\rm circ}
\ .
\end{eqnarray*}
Hence, the $N$-body simulations showing an intermediate orbit elongation for
maximum global potential tides \cite{mw87} are understood, since
when the cluster region is nearly homogeneous, the effective tide 
increases with increasing pericenter velocity, but not when the cluster density
profile decreases sharply as outside the core of the Modified Hubble model used
in the simulations.

\subsection{Mergers}

Because galaxies have their own internal energy, galaxy collisions are often
inelastic enough to lead to merging.
The timescale for merging may be estimated from a merging cross-section, again
as
$$
\tau_m = n \left\langle \sigma v \right\rangle^{-1}
\ .
$$
Using a numerical experimental cross-section \cite{rn79}, and integrating
over a Maxwellian velocity distribution, the
merger time can be written \cite{m92}
$$
\tau_m = {\rm Cst} \left [n r_g^2 v_g K(v_{\rm cl}/v_g)\right ]^{-1} \ ,
$$
where $n$ is the number density of galaxies, $r_g$ and $v_g$ are the galaxy
half-mass radius and internal velocity dispersion, respectively,
and the dimensionless merging efficiency $K$ is optimum for groups 
($v_{\rm cl} \simeq v_g$), while for clusters it falls off as $v_{\rm
cl}^3$ \cite{m92}. 
In groups as dense as compact groups \cite{h82} appear to be, merging
ought to be extremely efficient, and the relatively low fraction of ellipticals
indicates that chance alignments are contaminating the compact group
catalogs \cite{m92}. Despite their high velocity dispersions, rich clusters
are able to produce the right amount of mergers to produce elliptical
morphologies \cite{m92}, and moreover, merging is able to account \cite{m92}
for the
morphology-density \cite{pg84}
and morphology-radius \cite{wg91}
relations.

\section{Cluster evolution}

\subsection{Dynamical evolution}

The physical processes described in \S\ \ref{timescales} compete in the
evolution of the galaxy 
system.
For example, merging leads to increased merger cross-sections, hence to a
merging instability \cite{oh77}.
However, this instability is slowed
down by tidal processes which are usually thought to truncate galaxies of their
outlying particles which become unbound \cite{m87}.
Yet, if the merging cross-section is related to the galaxy half-mass
radius \cite{af80}, 
and since the tidal processes for galaxies on elongated
orbits or from collisions pump energy into the system, then the half-mass
radius of those particles that remain bound to the galaxy should increase.
The question remains whether the new half-mass radius is then greater or
smaller than
the old value, but this reviewer is not aware
of any numerical study that has addressed this question yet.

Similarly, if galaxies possess huge halos when they enter clusters (as one
can infer from Table \ref{tb:rta}), they
should feel strong dynamical friction and orbital decay.
But once they pass near the cluster core, these halos should be severely
stripped by the global tide of the cluster potential, after which the galaxy
is less massive and will be no longer subject to much dynamical friction. 

The interplay of the various dynamical processes and the difficulty in
analytical modeling of tidal effects render necessary to run numerical
$N$-body simulations to see how 
groups and clusters evolve. 
%
%
In collisionless particle simulations,
galaxies rapidly dissolve in clusters \cite{wdef87}, a problem known as {\sl
overmerging\/}, which is often ascribed to 
numerical effects.
In particular, two-body relaxation between galaxy particles and cluster
particles has been blamed for overmerging \cite{c94}.
Two analytical estimates differ on the importance of this effect.
Whereas, one \cite{vk95} finds a short timescale for the two-body relaxation
between galaxy particles and cluster particles:
$\tau_{\rm cl-g} = (v_g/v_{\rm cl})^2 \tau_{\rm
cl-cl} \simeq 0.1\, \tau_{\rm cl-cl}$, the other study \cite{mkl95} considers
the 
time $\tau = U / \dot U = U / (\dot N \Delta U_{\rm imp. apx.\/})$, where $U$
is the 
internal energy of the galaxy and $\dot N$ is the rate of collisions, and
finds a timescale typically longer than the Hubble time.
This last study \cite{mkl95} also checks that evaporation is not responsible
for overmerging.

Overmerging can also be ascribed to physical effects such as the tides from
the global potential of the cluster or dense group or two-body relaxation
between galaxies.
In semi-analytical calculations of the passage
through a cuspy ($\rho \sim r^{-2}$ density profile) primary potential, 
of a smooth secondary, also with a cuspy potential,
one finds \cite{gms94} that, for reasonably elongated orbits,
the tides of the primary
pump in
considerably more energy into the secondary
than its own binding energy.
Hence, secondaries such as galaxies dissolve at first passage.
It remains to be seen whether these secondaries are completely disrupted or
whether energy exchanges within it allow a dense core to survive.
Recent simulations \cite{mkl95} of multi-particle secondaries moving along
elongated 
orbits in a given isothermal potential show that tidal disruption is
relatively slow, although it is not clear whether their ``isothermal''
primary is the singular $\rho \sim r^{-2}$ model or the non-singular 
isothermal (that thus has an homogeneous core), for which one indeed expects
softer tidal effects. 

In any event, overmerging is much less evident in simulations where gas is
included \cite{kw93b,esd94},
presumably because the gas sinks to the bottom of the galaxy
potential wells and deepens these wells, which thus avoid merging with one
another. 
The latter study emphasizes the importance of star formation in their gaseous
galaxies in reducing overmerging.

Dense groups of galaxies witness rapid merging and coalesce into a single
elliptical galaxy \cite{ccs81,b85b,m87,b89}.
A detailed comparison of the results on groups \cite{m90}
showed that the
different numerical studies of groups produced comparable rates of merging.
The differences arise in part from the initial conditions and from the fact
that the merger rate in dense groups is decreased when most of the dark
matter is distributed in a common envelope, rather than in individual galaxy
halos \cite{cstv82,b85b,m87,bcl93}.
Indeed, with large individual halos, galaxy merging is direct, while the
presence of an important intergalactic background causes galaxies to
dissipate orbital energy by dynamical friction, suffer orbital decay and
finally merge together at the group center \cite{m87}.
It may be that overmerging is occurring in all these dense group simulations,
and that with the inclusion of gas dynamics, the galaxies in dense groups may
survive longer.


\subsection{Cosmological evolution}

Spherical top-hat cosmology (\S\ \ref{th-density}), can be applied to a
galaxy system 
bathing in an empty or uniform universe.
%
%
An homogeneous isolated system should see its size evolve as shown in Figure
1a. 
\begin{figure}[!ht]
\begin{center}
\psfig{figure=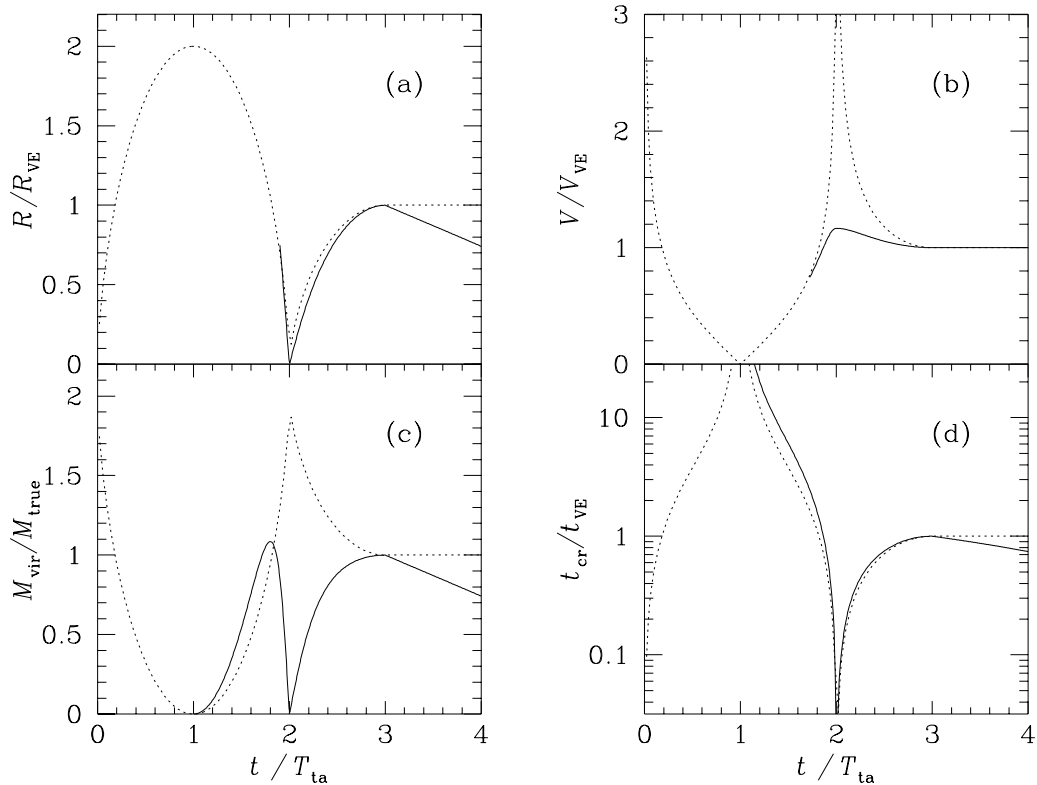,width=15cm,angle=-90}
\caption{Time evolution of bias in
observed 
radius (a), velocity dispersion (b), mass (c), and crossing time (d), 
relative to virial equilibrium (VE), where
$T_{\rm ta}$ is the turnaround time.
The {\it dotted curves\/} show the evolution for point-masses, while the {\it
full curves\/} show the effects of softened potentials and orbital energy
dissipation by dynamical friction (starting at $t = 3\,T_{\rm ta}$).}
\end{center}
\label{fig1}
\end{figure}
It first follows the Hubble expansion, then decouples from this expansion and
turns around, 
collapses and subsequently virializes.
If a system is in dynamical equilibrium, one can
apply the virial theorem, and so derive a {\sl virial mass\/} $M_{\rm vir} =
R V^2/ 
G$.
While the time evolution of the radius $R$, velocity dispersion $V$, ratio of
virial mass to true mass $M_{\rm vir}/M$ and ratio of crossing time to
running time 
$t_{\rm cr}/t = R/(Vt)$  parameters have been studied 
before \cite{ggmmr88} (though simple $N$-body simulations), and an analytical
version is shown in Figures 1b, 1c, 
and 1d, one gains considerable insight in plotting the evolution of
$M_{\rm vir}/M$ versus $t_{\rm cr}/t$ as is done in Figure 2a.
The dotted track is for groups made of point mass galaxies, while the solid
track is for extended galaxies, which reach a terminal velocity at group
collapse (because the smoothed potential is flat at the center), and after
virialization, dissipate their orbital energy by dynamical 
friction against their common massive halo (merged from their individual halos
after group collapse).

\begin{figure}[!ht]
\begin{center}
\psfig{figure=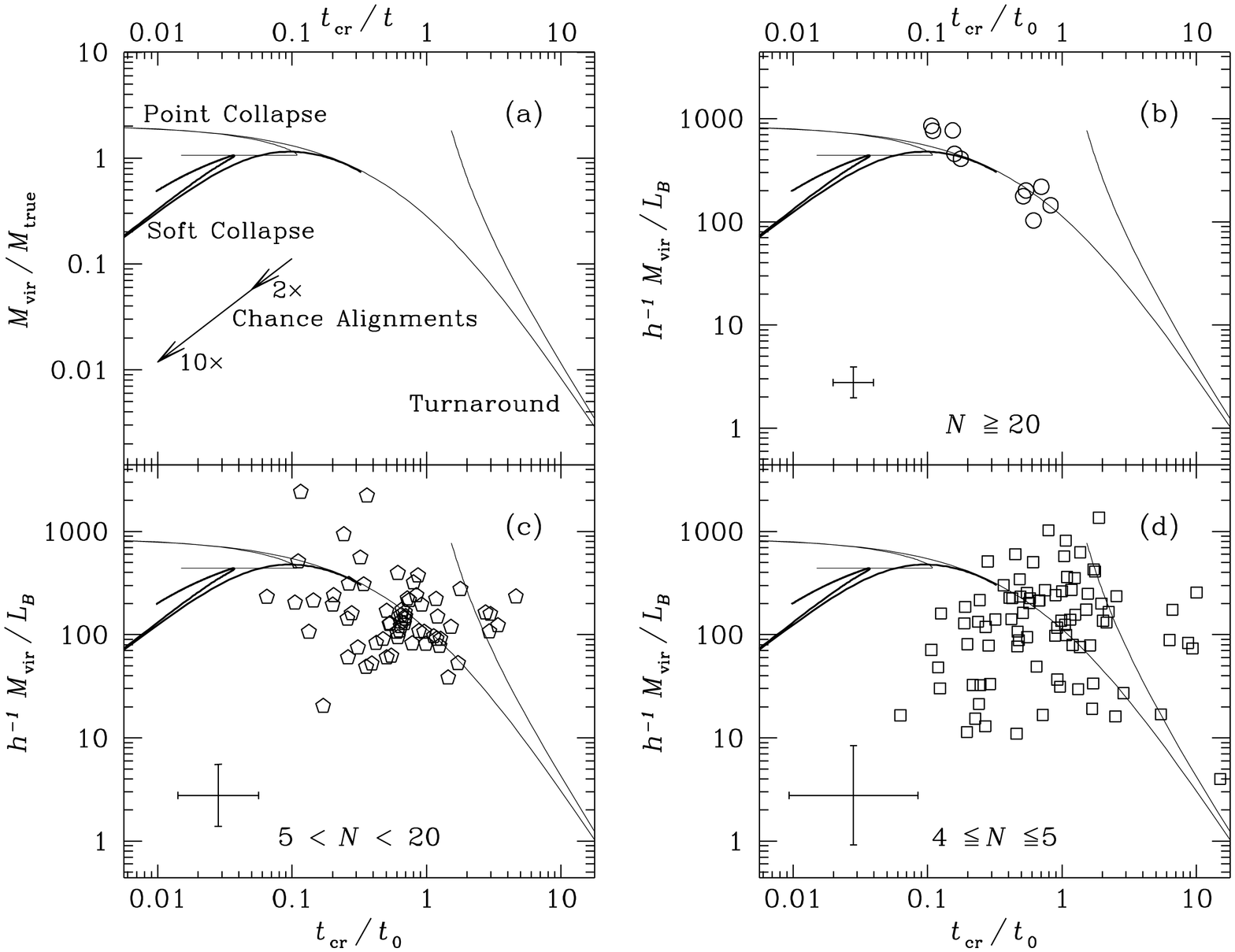,width=15cm,angle=-90}
\caption{Mass assuming dynamical equilibrium, scaled to total
mass (a) or total blue luminosity (b, c, and d), versus crossing time (in units
of the age of the Universe for $\Omega = 1$, while for $\Omega = 0.2$ the
points should be displaced to the left by 0.1 decade).
The {\it polygons\/} (b, c, and d) represent the loose groups.
The {\it thin curves\/} are the theoretical point-mass evolutionary tracks,
while the {\it thick curves\/} are the same for softened potentials and
allowing for orbital energy dissipation after virialization.
In (b, c, and d), these curves are scaled to mass-to-light ratios assuming that
all groups have a true $M/L = 440\,h$. The typical
error bar for each data point is shown.}
\end{center}
\label{fig2}
\end{figure}

The solid track in Figure 2a can be considered as a {\sl fundamental track\/}
that galaxy systems should follow, similar in concept to a Hertzprung-Russell
diagram for cosmological galaxy systems.
Unfortunately, true masses of galaxy systems are not known.
To compare with observed parameters, we must make an assumption on the true
mass, and the simplest one is to assume that the true mass follows light,
{\it i.e.\/,} $M/L = \rm cst$.
In Figures 2b, c, and d, we plot the observable parameters, $M_{\rm vir}/L_B$
vs. 
$H_0 t_{\rm cr}$, for loose groups \cite{gcf92} of different multiplicities,
and superpose the 
theoretical evolutionary track, adjusting the $y$-axis with the high
multiplicity groups of Figure 2b, while the $x$-axis scaling is imposed by
theory. 

The high-multiplicity groups fit the theoretical tracks very well.
A one proceeds to lower multiplicities, the statistical noise in the
mass-to-light ratio and crossing time estimates increases, but so does the
probability for chance alignments, which make the groups appear smaller while
conserving on the average their velocity dispersion.
Although precise assignments of group cosmo-dynamical states is difficult
because of 
statistical noise, one can nevertheless get a handle on which groups are
unbound (above theoretical track), which are still in 
their expansion phase (upper-right handle of track), which are near
turnaround (lower-right handles of track), which are collapsing
(central handle), which are near maximum collapse (first lower-left
handle), and those that are virialized (second lower-left handle).
The theoretical fundamental track in observable space $M_{\rm vir}/L$
vs. $t_{\rm cr}/t_0$ thus 
represents a slice through the {\sl fundamental 
surface\/} (which is curved) of groups, where the third axis can be the scale
of the system, such as its total optical
luminosity.
In contrast, an empirical fundamental {\it plane\/} has been
reported \cite{bbfn95} for loose groups, similar to that of elliptical
galaxies and clusters (see \S\ \ref{fp}).

The true $M/L$ is obtained by extrapolating to the early virialized state
(before dissipation of orbital energy, which occurs at nearly constant velocity
dispersion since the common halo should have near constant circular velocity).
The loose groups \cite{gcf92} then have $M_{\rm true}/L = 440\,h$, much higher
than the median $M/L = 130\,h$, for the groups of $N \geq 4$ members (the mass
estimate used here is the median of the non-weighted virial, weighted virial,
and projected masses).
In other words, {\it the mass-to-light ratios of groups are severely
underestimated 
because most groups are still relatively near their turnaround phase.\/}
This points to $\Omega \simeq 0.3$ obtained by extrapolating 
the galaxy luminosity function \cite{lpem92} to $(M/L)_{\rm closure} = 1560\,h$.

In any event, {\it no
groups in the loose group catalog has yet completed its collapse\/}, not even
the Virgo cluster included in the catalog, whose outer members are still
collapsing onto the virialized core.
The dynamical youth of loose groups suggests a high value (close to unity)
for $\Omega_0$.
Details of this analysis are still in preparation.

The fundamental tracks for groups can been computed in a hierarchical binary
approximation, with a galaxy merging with an already merged pair.
The precise position of the track depends slightly on the mass ratio of the
merging pair.
Any system can be represented at a given hierarchy.
For example, the NGC 2300 group, in which X-rays from the intergalactic gas
were first discovered \cite{mdmb93}, is a loose triplet that has just turned
around from its maximum Hubble expansion, while the tight binary around 
which the X-rays are roughly centered appears on the evolved part of the
fundamental track (late collapse or rebound).
The fundamental surface analysis can be easily applied to pairs,
but the error bars in the position of a pair on the diagram are relatively
important, and a detailed statistical analysis is required.
Conversely, clusters suffer little from statistical uncertainties of the
observed parameters, but when a
subcluster is merging into a cluster, 
the fundamental surface analysis on the primary cluster may break down as the
secondary can exert non-negligible tidal stresses that alter the collapse of
the primary.

In the previrialization scenario, structures acquire angular momentum at
maximum expansion, and do not fully collapse to reach a rapid virial
equilibrium.
The fundamental track, should then be slightly altered, as it will
follow the standard track at early epochs (and high observed crossing times,
to the right of Fig. 2), and up to $t_{\rm cr}/t \simeq 0.03$, then slightly
rebound to the virialized horizontal wing (the thin horizontal track to the
left of Fig. 2). 
If the previrialized system coalesces by dynamical friction, its track will
then follow the soft collapse track, and its overall behavior will be
similar to the standard track.
However, a coalescence is somewhat akin
to a collapse, but slower. Hence, the the decrease of the track to the lower
left of the diagram is delayed with respect to the standard track.

In Figure 3 are shown the locations of compact groups \cite{hmhp92}
in the fundamental track diagram.
The groups on the upper left of the diagram are high velocity dispersion
groups close to the fundamental track.
Conversely, the groups on the lower right of the diagram are low velocity
dispersion groups too far from the fundamental track to be explained by
standard evolution. Instead, they are best explained as arising from chance
alignments of galaxies along the line of sight, either within larger loose
groups \cite{m86} that are still collapsing, or within  long ($\sim 7 \,
h^{-1} \, \rm Mpc$) filaments
 \cite{hkw95}, often seen in cosmological simulations, and some of which may
be expanding (the spherical equivalent being the upper right track).
Thus, the higher velocity dispersion half of this sample of compact groups
are real 3-D dense systems, either terminating their collapse or rebounding
from it, or in their final coalescence stage, while the lower velocity
dispersion half are caused by chance alignments.
If previrialization is not followed by rapid coalescence, then
as much as 90\% of the compact groups would lie too far from the fundamental
track to be real and would be ascribed to chance alignments.
This is inconsistent with the detection of X rays, probing the potential of
truly dense groups, in roughly half of these compact
groups \cite{mdmb95} (typically the high velocity dispersion ones
\cite{mdmb95,mh96}).
If coalescence does occur after previrialization, the data of Figure 3 is not
good enough to distinguish between the two scenarios, but a richer compact
group catalog,
or the use of X-rays to infer in a more secure fashion the velocity
group dispersion should help settle between them.

\begin{figure}[!ht]
\begin{center}
\psfig{figure=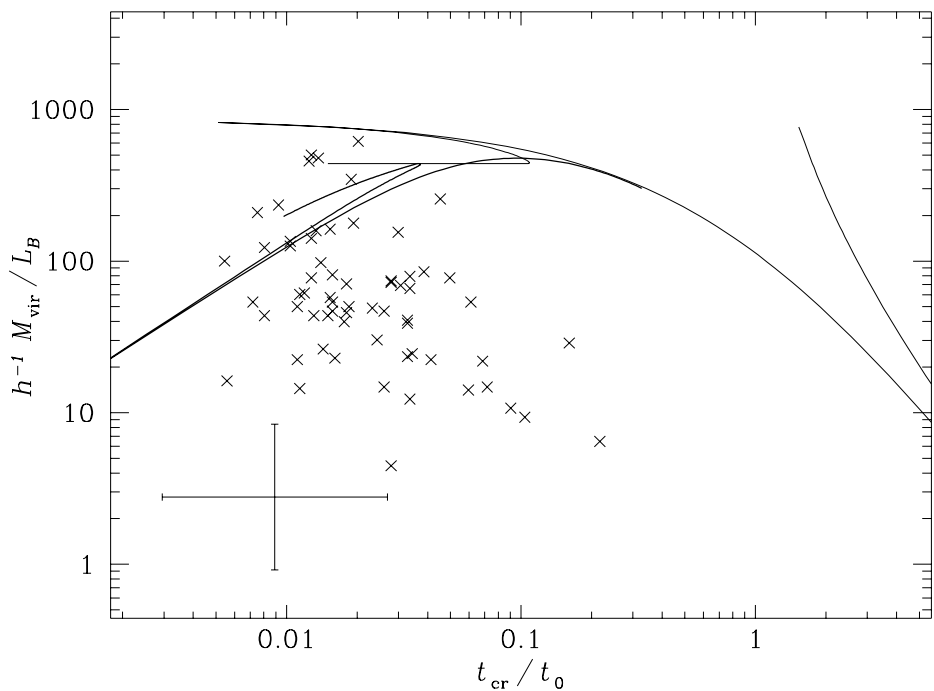,width=13cm,angle=-90}
\caption{Same as Figures 2bcd for compact groups
({\it crosses\/}).}
\end{center}
\label{fig3}
\end{figure}



\begin{thebibliography}{9}
\parskip=-5pt
\bibitem{a58} G.O. Abell {\it Ap. J. Supp. Ser.\/} {\bf 3} (1958), 211.
\bibitem{h82} P. Hickson {\it Ap. J.\/} {\bf 255} (1982), 382.
\bibitem{t76} E.L. Turner {\it Ap. J.\/} {\bf 208} (1976), 20.
\bibitem{b75} N.A. Bahcall {\it Ap. J.\/} {\bf 198} (1975), 249.
\bibitem{y74} A. Yahil {\it Ap. J.\/} {\bf 191} (1974), 623.
\bibitem{bt86} T. Beers and J.L. Tonry {\it Ap. J.\/} {\bf 300} (1986), 557.
\bibitem{mg94} D. Merritt and K. Gebhardt (1994)
in XIVth Moriond Astrophysics Meeting {\it Clusters of Galaxies\/},
ed. F. Durret, A. Mazure and J. Tran Thanh Van (Fronti\`eres, Gif-sur-Yvette,
1994), p. 11, astro-ph/9510091.
\bibitem{t87} R.B. Tully {\it Ap. J.\/} {\bf 321} (1987), 280.
\bibitem{wm89} D.G. Walke and G.A. Mamon {\it Astr. Ap.\/} {\bf 225} (1989).
291. 
\bibitem{hnhm84} P. Hickson, Z. Ninkov, J.P. Huchra and G.A. Mamon in
{\it Clusters and Groups of Galaxies\/}, ed.  F. Mardirossian, G. Giuricin,
and M. Mezzetti (Reidel, Dordrecht, 1984), p. 367.
\bibitem{mdgms96} M.-L. Montoya, R. Dom\'{\i}nguez-Tenreiro,
G. Gonz\'alez-Casado, 
G.A. Mamon and E. Salavador-Sol\'e in 
{\it From Galaxies to Galaxy Systems\/} also {\it Ap. Lett. Comm.\/} (1996),
in press.
\bibitem{hmhp92} P. Hickson, C. Mendes de Oliveira, J.P. Huchra and
G.G.C. Palumbo {\it 
Ap. J.\/} {\bf 399} (1992), 353. 
\bibitem{bm82} J. Binney and G.A. Mamon {\it M.N.R.A.S.\/} {\bf 200} (1982),
361. 
\bibitem{flg80} D.M. Fabricant, M. Lecar and P. Gorenstein {\it Ap. J.\/}
{\bf 241} (1980), 552.
\bibitem{c74} A. Cavaliere {\it Mem. Soc. Astron. It.\/} {\bf 44} (1974), 571.
\bibitem{h85} M.J. Henriksen {\it Ph.D. Thesis, Univ. of Maryland\/} (1985).
\bibitem{h89} J.P. Hughes {\it Ap. J.\/} {\bf 337} (1989), 21.
\bibitem{gdll92} D. Gerbal, F. Durret, M. Lachi\`eze-Rey and G. Lima-Neto
{\it Astr. Ap.\/} {\bf 253} (1992), 77.
\bibitem{dglls94} F. Durret, D. Gerbal, M. Lachi\`eze-Rey, G. Lima-Neto and
R. Sadat {\it Astr. Ap.\/} {\bf 287} (1994), 733.
\bibitem{bk75} 
R. Bourassa and R. Kantowski {\it Ap. J.\/} {\bf 195} (1975), 13.
\bibitem{gn88} S.A. Grossman and R. Narayan {\it Ap. J.\/} {\bf 324}
(1988), L77. 
\bibitem{sfmp87} G. Soucail, B. Fort, Y. Mellier and J.-P. Picat {\it
Astr. Ap.\/} {\bf 172} (1987), L14.
\bibitem{lp86} R. Lynds and V. Petrosian {\it Bull. Am. Astr. Soc.} {\bf 18}
(1986), 1014.
\bibitem{ks93} N. Kaiser and G. Squires {\it Ap. J.\/} {\bf 404} (1993), 441.
\bibitem{tf95} A.J. Tyson and P. Fischer {\it Ap. J.\/} {\bf 446}
(1995), L55, astro-ph/9503119.
\bibitem{skbfwnb95} G. Squires, N. Kaiser, A. Babul, G. Fahlman, D. Woods,
D.M. Neumann and H. B\"ohringer {\it Ap. J.\/}, submitted (1995),
astro-ph/9507008.
\bibitem{lm94} A. Loeb and S. Mao {\it Ap. J.\/} {\bf 435} (1994), L109,
astro-ph/9406030.
\bibitem{mb95} J. Miralda-Escud\'e and A. Babul {\it Ap. J.\/} {\bf 449}
(1995), 18, astro-ph/9405063.
\bibitem{ans64} G.O. Abell, J. Neyman and E. Scott {\it Astr. J.\/} {\bf 69}
(1964), 529.
\bibitem{b77} F.W. Baier {\it Astr. Nach.\/} {\bf 298} (1977), 151.
\bibitem{gb82} M.J. Geller and T.C. Beers {\it P.A.S.P.\/} {\bf 94} (1982),
421.
\bibitem{bgbh83} G. Bothun, M.J. Geller, T.C. Beers and J.P. Huchra {\it
Ap. J.\/} {\bf 268} (1983), 47.
\bibitem{jf92} C. Jones and W. Forman in {\it Clusters and Superclusters of
Galaxies\/}, ed. A.C. Fabian (Kluwer, Dordrecht, 1992), p. 49.
\bibitem{ds88} A. Dressler and S.A. Shectman {\it Astr. J.\/} {\bf 95}
(1988), 985. 
\bibitem{b84} F.W. Baier {\it Astr. Nach.\/} {\bf 305} (1984), 175.
\bibitem{sgs93} E. Salvador-Sol\'e, G. Gonz\'alez-Casado and J.M. Solanes
{\it Ap. J.\/} {\bf 410} (1993), 1.
\bibitem{ebggmmm94} E. Escalera, A. Biviano, M. Girardi, G. Giuricin,
F. Mardirossian, A. Mazure and M. Mezzetti {\it Ap. J.\/} {\bf 423} (1994),
539, astro-ph/9309057.
\bibitem{hr88} P. Hickson and H.J. Rood {\it Ap. J.\/} {\bf 331}
(1988) L69.
\bibitem{smcb92} R. Schaeffer, S. Maurogordato, A. Cappi and F. Bernardeau
{\it M.N.R.A.S.\/} {\bf 263} (1992), L21.
\bibitem{kg82} S.M. Kent, J.E. Gunn {\it Astr. J.\/} {\bf 87} (1982), 945.
\bibitem{b82} J. Binney {\it M.N.R.A.S.\/} {\bf 200} (1982), 951.
\bibitem{hk95} R. den Hartog, P. Katgert {\it M.N.R.A.S.\/} (1995), submitted.
\bibitem{gg72} J.E. Gunn, J.R. Gott {\it Ap. J.\/} {\bf 176} (1972), 1.
\bibitem{m92} G.A. Mamon {\it Ap. J.\/} {\bf 401} (1992), L3.
\bibitem{b85} E. Bertschinger {\it Ap. J. Supp. Ser.\/} {\bf 58} (1985), 39.
\bibitem{cl95} S. Cole and C. Lacey {\it M.N.R.A.S.\/}, submitted (1995),
astro-ph/9510147.
\bibitem{gh83} M.J. Geller and J.P Huchra {\it Ap. J. Suppl.\/} {bf 52}
(1983), 61.
\bibitem{rgh89} M. Ramella, M.J. Geller and J.P Huchra {\it Ap. J.\/} {bf 344}
(1989), 57.
\bibitem{m94} G.A. Mamon in XIVth Moriond Mtg. {\it Clusters of Galaxies\/},
ed. F. Durret, A. Mazure and J. Tran Van Thanh (Gif-sur-Yvette, Fronti\`eres,
1994), 291, astro-ph/9406043.
\bibitem{g75} J.R. Gott {\it Ap. J.\/} {\bf 201} (1975), 296.
\bibitem{fg84} J.A. Fillmore, P. Goldreich {\it Ap. J.\/} {\bf 281} (1984), 1.
\bibitem{mabp95} F. Moutarde, J.-M. Alimi, F.R. Bouchet, R. Pellat {\it
Ap. J.\/} {\bf 441} (1995), 10.
\bibitem{g77} J.E. Gunn {\it Ap. J.\/} {\bf 218} (1977), 592.
\bibitem{zqs88} W.H. Zurek, P.J. Quinn, J.K. Salmon {\it Ap. J.\/} {\bf 330}
(1988), 519.
\bibitem{cer94} M.M. Crone, A.E. Evrard, D.O. Richstone {\it Ap. J.\/} {\bf
434} (1994), 402, astro-ph/9404030.
\bibitem{hs85} Y. Hoffman, J. Shaham {\it Ap. J.\/} {\bf 297} (1985), 16.
\bibitem{efwd88} G. Efstathiou, C.S. Frenk, S.D.M. White, M. Davis {\it
M.N.R.A.S.\/} {\bf 235} (1988), 715.
\bibitem{dc91} J. Dubinski and R.G. Carlberg {\it Ap. J.\/} {\bf 378} (1991),
496.
\bibitem{nfw95} J.F. Navarro, C.S. Frenk and S.D.M. White {\it M.N.R.A.S.\/}
{\bf 275} (1995), 720, astro-ph/9408069.
\bibitem{gr75} J.R. Gott and M. Rees {\it Astr. Ap.\/} {\bf 45} (1975), 365.
\bibitem{rlt92} D. Richstone, A. Loeb and E.L. Turner {\it Ap. J.\/} {\bf
393} (1992), 477.
\bibitem{ps74} W.H. Press and P. Schechter {\it Ap. J.\/} {\bf 187} (1974),
425. 
\bibitem{lc93} C. Lacey and S. Cole {\it M.N.R.A.S.\/} {\bf 262} (1993), 627.
\bibitem{kw93} G. Kauffmann and S.D.M. White {\it M.N.R.A.S.\/} {\bf 261},
921.
\bibitem{gms94} G. Gonz\'alez-Casado, G.A. Mamon and E. Salvador-Sol\'e {\it
Ap. J.\/} {\bf 433} (1994), L61, astro-ph/9406066.
\bibitem{gss93} G. Gonz\'alez-Casado, J.M. Solanes and E. Salvador-Sol\'e {\it
Ap. J.\/} {\bf 410} (1994), 15.
\bibitem{emfg93} A.E. Evrard, J.J. Mohr, D.J. Fabricant and M.J. Geller {\it
Ap. J.\/} {\bf 419} (1993), L9.
\bibitem{mefg95} J.J. Mohr, A.E. Evrard, D.J. Fabricant and M.J. Geller {\it
Ap. J.\/} {\bf 447} (1995), 8, astro-ph/9501011.
\bibitem{dp77} M. Davis and P.J.E. Peebles {\it Ap. J. Supp.\/} {\bf 34}
(1977), 425.
\bibitem{p90} P.J.E. Peebles {\it Ap. J.\/} {\bf 365} (1990), 27.
\bibitem{p80} P.J.E. Peebles {\it Large Scale Structure of the Universe\/},
Princeton, Princeton University Press (1980), chap. 19.
\bibitem{ec92} A.E. Evrard and M.M. Crone {\it Ap. J.\/} {\bf 394}, L1.
\bibitem{ljbh96} E.L. {\L}okas, R. Juszkiewicz, F.R. Bouchet and E. Hivon {\it
Ap. J.\/} in press (1996), astro-ph/9508032.
\bibitem{wssd95} M. White, D. Scott, J. Silk and M. Davis {\it M.N.R.A.S.\/},
in press (1995), astro-ph/9508009.
\bibitem{m93} G.A. Mamon in {\it N-Body Problems and Gravitational
Dynamics\/}. ed. F. Combes and E. Athanassoula, (Meudon, Obs. de Paris, 1993),
188, astro-ph/9308032.
\bibitem{c42} S. Chandrasekhar {\it Principles of Stellar Dynamics\/} (1942),
New York, Dover.
\bibitem{gs86} V.G. Gurzadyan and G.K. Savvidy {\it Astr. Ap.\/} {\bf 160}
(1986), 203.
\bibitem{lb67} D. Lynden-Bell {\it M.N.R.A.S.\/} {\bf 136} (1967), 101.
\bibitem{t77} A. Toomre in {\it The Evolution of Galaxies and Stellar
Populations\/}, ed. B.M. Tinsley and R.B. Larson (New Haven, Yale
Univ. Press, 1977), p. 401.
\bibitem{a38} V.A. Ambartsumian {\it Ann. Leningrad State Univ.\/} {\bf 22}
(1938), 19, translated in IAU Symp. 113 {\it Dynamics of Star Clusters\/},
ed. J. Goodman and P. Hut (Dordrecht, Reidel, 1985), p. 521.
\bibitem{c43} S. Chandrasekhar {\it Ap. J.\/} {\bf 97} (1943), 255.
\bibitem{lt83} D.N.C. Lin and S.D. Tremaine {\it Ap. J.\/} {\bf 264} (1983),
364.
\bibitem{pc92} P. Prugniel and F. Combes {\it Astr. Ap.\/} {\bf 259} (1992),
25.
\bibitem{cgm95} R. Chan, G.A. Mamon and D. Gerbal {\it Astr. Ap.} (1995),
submitted.
\bibitem{m85} D. Merritt {\it Ap. J.\/} {\bf 289} (1985), 18.
\bibitem{r75} D.O. Richstone {\it Ap. J.\/} {\bf 200} (1975), 535.
\bibitem{dls80} A. Dekel, M. Lecar and J. Shaham {\it Ap. J.\/} {\bf 241}
(180), 946.
\bibitem{osc72} J.P. Ostriker, L. Spitzer and R.A. Chevalier {\it Ap. J.\/}
{\bf 176} (1972), L51.
\bibitem{s58} L. Spitzer {\it Ap. J.\/} {\bf 127} (1958), 17.
\bibitem{m87} G.A. Mamon {\it Ap. J.\/} {\bf 321} (1987), 622.
\bibitem{ar88} A.J. Allen and D.O. Richstone {\it Ap. J.\/} {\bf 325} (1988),
583.
\bibitem{mw87} D. Merritt and S.D.M. White in IAU Symp. 117 {\it Dark Matter in
the Universe\/}, ed. J. Kormendy \& G.R. Knapp (Dordrecht, Reidel, 1987) 283.
\bibitem{rn79} N. Roos and C.A. Norman {\it Astr. Ap.\/} {\bf 95} (1979), 349.
\bibitem{pg84} M. Postman and M.J. Geller {\it Ap. J.\/} {\bf 281} (1984), 95.
\bibitem{wg91} B.C. Whitmore and D. Gilmore {\it Ap. J.\/} {\bf 367} (1991),
64.
\bibitem{oh77} J.P. Ostriker and M. Hausman {\it Ap. J.\/} {\bf 217} (1977),
L125.
\bibitem{af80} S. Aarseth and S.M. Fall {\it Ap. J.\/} {\bf 236} (1980), 43.
\bibitem{wdef87} S.D.M. White, M. Davis, G. Efstathiou and C.S. Frenk {\it
Nature\/} {\bf 330} (1987), 451.
\bibitem{c94} R.G. Carlberg {\it Ap. J.\/} {\bf 433} (1994), 468.
\bibitem{vk95} E. van Kampen {\it M.N.R.A.S.\/} {\bf 273} (1995), 295.
\bibitem{mkl95} B. Moore, N. Katz and G. Lake {\it Ap. J. (Lett.)\/},
submitted (1995), astro-ph/9503088.
\bibitem{kw93b} N. Katz and S.D.M. White {\it Ap. J.\/} {\bf 412} (1993), 455.
\bibitem{esd94} A.E. Evrard, F. Summers and M. Davis {\it Ap. J.\/} {\bf
422} (1994), 11.
%
\bibitem{ccs81} P. Carnevali, A. Cavaliere and P. Santangelo {\it Ap. J.\/}
{\bf 249} (1981), 449.
\bibitem{b85b} J. Barnes {\it M.N.R.A.S.\/} {\bf 215} (1985), 517.
\bibitem{b89} J. Barnes {\it Nature\/} {\bf 338} (1989), 123.
\bibitem{m90} G.A. Mamon in IAU Coll. 124 {\it Paired and Interacting
Galaxies\/}, ed. J.W. Sulentic \& W.C. Keel (Washington, NASA, 1990), 609.
\bibitem{cstv82} A. Cavaliere, P. Santangelo, G. Tarquini and N. Vittorio in
{\it Clustering in the Universe\/}, ed. D. Gerbal \& A. Mazure
(Gif-sur-Yvette, Fronti\`eres, 1982) p. 25.
\bibitem{bcl93} P.W. Bode, H.N. Cohn and P.M. Lugger {\it Ap. J.\/} {\bf 416}
(1993), 17.
\bibitem{ggmmr88} G. Giuricin, P. Gondolo, F. Mardirossian, M. Mezzetti and
M. Ramella {\it Astr. Ap.\/} {\bf 199} (1988), 85.
\bibitem{gcf92} E. Gourgoulhon, P. Chamaraux and P. Fouqu\'e {\it
Astr. Ap.\/} {\bf 255} (1992), 69.
\bibitem{bbfn95} D. Burstein, R. Bender, S.M. Faber and R. Nolthenius {\it
Ap. Lett. Comm.\/} {\bf 31} (1995), 95.
\bibitem{lpem92} J. Loveday, B.A. Peterson, G. Efstathiou and S.J. Maddox
{\it Ap. J.\/} {\bf 390} (1992), 338.
\bibitem{mdmb93} J.S. Mulchaey, D. Davis, R.F. Mushotzky and D. Burstein
{\it Ap. J.\/} {\bf 404} (1993), L9.
\bibitem{m86} G.A. Mamon {\it Ap. J.\/} {\bf 307} (1986), 426.
\bibitem{hkw95} L. Hernquist, N. Katz and D.H. Weinberg {\it ApJ\/} {\bf 442}
(1995), 57, astro-ph/9407059.
\bibitem{mdmb95} J.S. Mulchaey, D. Davis, R.F. Mushotzky and D. Burstein
{\it Ap. J.\/} in press (1995).
\bibitem{mh96} G.A. Mamon and M.J. Henriksen, in preparation.
\end{thebibliography}
\end{document}